\documentclass[traditabstract]{aa} 
\usepackage{txfonts,epsfig,graphicx,natbib,url,twoopt}

\newcommand{\thetabold}{\mbox{\boldmath$\theta$}}
\newcommand{\Thetabold}{\mbox{\boldmath$\Theta$}}

\newcommand{\Omegabold}{\mbox{\boldmath$\Omega$}}

\begin{document}

   \title{Coronal loop physical parameters from the analysis of multiple observed transverse oscillations}

   \author{A. Asensio Ramos \and I. Arregui
          }

   \institute{Instituto de Astrof\'\i sica de Canarias,
              38205, La Laguna, Tenerife, Spain; \email{aasensio@iac.es}
            \and
Departamento de Astrof\'{\i}sica, Universidad de La Laguna, E-38205 La Laguna, Tenerife, Spain
             }

  \abstract{The analysis of quickly damped transverse oscillations of solar coronal loops using
  magneto-hydrodynamic seismology allow us to infer physical parameters that are difficult to measure otherwise.
  Under the assumption that such damped oscillations are due to the resonant conversion of global modes into
  Alfv\'en oscillations of the tube surface, we carry out a global seismological analysis of a large set of coronal 
  loops. A Bayesian hierarchical method is used to obtain distributions for coronal loop physical parameters by means of a global 
  analysis of a large number of observations. The resulting distributions summarise global information and  constitute data-favoured 
  information that can be used for the inversion of individual events.
  The results strongly suggest that internal Alfv\'en travel times along the loop are larger than 100 s and smaller than 540 s
  with 95\% probability. Likewise, the density contrast between the loop interior and the surrounding is 
  larger than 2.3 and below 6.9 with 95\% probability.}

   \authorrunning{Asensio Ramos \& Arregui}
   \maketitle
%

\section{Introduction}

The discovery of quickly damped transverse oscillations of solar coronal loops was
first reported by \cite{aschwanden99} and \cite{nakariakov99} using Transition Region 
and Coronal Explorer (TRACE) observations. The phenomenon was interpreted in terms of the 
standing linear magnetohydrodynamic (MHD) kink mode of a magnetic flux tube in its fundamental 
harmonic. The cause of the quick damping of the oscillations has been attributed to the resonant 
conversion of global motions into localised Alfv\'en oscillations at the tube boundary, 
because of the transverse inhomogeneity of the medium \citep{goossens02}. The essence of the kink 
mode has been found to be of mixed nature \cite{goossens09} with a dominant Alfv\'enic character 
\citep{goossens12a}. Recent reviews on theoretical aspects of MHD kink waves can be found in 
\cite{ruderman09} and \cite{goossens11}. Observational properties of transverse coronal loop oscillations
are presented and discussed by \cite{schrijver02} and \cite{aschwanden02}.

MHD seismology \citep{uchida70,roberts84} uses inversion techniques to infer
difficult to measure physical parameters combining theory and observations of
MHD waves. Coronal seismology applications using transverse loop oscillation
have been successful in determining parameters such as the magnetic field
strength \citep{Nakariakov01}, the Alfv\'en speed
\citep{arregui07,goossens08,goossens12b}, the transversal density structuring
\citep{verwichte06}, or the coronal density scale height
\citep{AAG05,Verth08,Arregui13}.

In a previous paper \citep{arregui_asensio11}, we pursued the Bayesian analysis of individual coronal loops
with the aim of inferring their fundamental parameters. In the Bayesian framework the inference is given by 
a distribution, the so-called posterior probability distribution, that is a combination of how well the 
observed data are predicted by the model, the likelihood, and our state of knowledge on the unknowns before 
considering the data, given by the priors. In that analysis, we demonstrated that the
inferred values of the Alfv\'en travel time are robust. Additionally, a Bayesian analysis is able to
give some information (at least put some constraints) on the transverse inhomogeneity. A result of the
work was the fact that the density contrast between the coronal loop and the ambient medium is poorly
constrained by the observations. The posterior distribution changes when different prior distributions are
used for this parameter. However, if an independent measurement of the density contrast is available,
we demonstrated that the three parameters can be accurately inferred from the period and damping
time of the coronal loop oscillations.

These previous studies have focused on the inversion of physical parameters
using measured wave properties for particular events, on a one-by-one basis,
thus obtaining estimates for the local properties of the plasma for each
particular event. Also, in most of the studies, the wave properties that are
used as input in the inversions consist of quantities that are obtained upon
manipulation of parameters obtained at the primary
stage of the data analysis. Some examples are the use of the period and damping
of the oscillations that are obtained after a fitting of the measured time
evolution of the displacement in a sequence of imaging observations, or the
phase speed of propagating waves derived from time-distance diagrams for
propagating waves.

This work represents a substantial step forward along two lines. First,
we go here to a more fundamental level than \cite{arregui_asensio11} and
use the displacement curves themselves, as measured by \cite{aschwanden02},
instead of period and damping times. The reason is that the assumption of a
gaussian likelihood function is more appropriate for the displacement curves
than for the derived quantities. The noise statistics in the derived quantities
is very complicated to obtain, given the non-trivial manipulations that are needed to
obtain them.

Second, a key issue in Bayesian parameter inference is the use of prior information that accounts for our
state of knowledge on the unknowns, before considering the data. This prior
knowledge is usually constructed on the basis of informed guesses about, e.g.,
the values and ranges of variation of physical parameters, physical constraints
imposed by the model, etc. In this work, we compute these priors using
observational information and obtained from the global analysis of a number of
observed events. This is done by performing a fully consistent analysis of a
large number of observations using a hierarchical Bayesian framework. In the
same way as directly measured properties of
transverse loop oscillations can be summarised by performing histograms, from
which quantities such as the mean, median or standard deviation can be
obtained, the Bayesian hierarchical framework enable us to obtain similar
information for the physical parameters that cannot be directly measured and
need thus to be inferred. As a result, data-favoured distributions for the
unknown parameters are obtained. They can then be used to construct priors based
on our current observational information of coronal loops.
 
The layout of the paper is as follows. Section 2 presents our inference approach
based on a Bayesian hierarchical model and how this is applied to the
observations. The sampling of the posterior and the marginalization is
also discussed in this section. Section 3 presents the final results of
the paper and we end with the conclusions in Section 4.

\section{Hierarchical modeling of coronal loops}
If one has direct observational access to a given physical quantity (for instance, brightness, displacement,
etc.), obtaining an estimation of the probability density for this quantity is easily
achieved by just counting events in bins. If observational uncertainties can shift events from one 
bin to another, it is possible to use Bayesian schemes to take this and other effects into
account \cite[e.g., the extreme deconvolution technique of][and references therein]{bovy_extremedeconv11}. However, when
one is interested in a quantity that cannot be directly observed but has to be inferred from
observations, the situation is not so straightforward. This is exactly the problem we have
in our case, because we do not have direct access to the physical parameters of the oscillating coronal loops. We propose a Bayesian hierarchical scheme to solve this
problem. It can be considered as an efficient way to estimate a probability density of an unobserved quantity,
obtained from many observations of quantities that are non-linearly related to the one of interest.

In summary, in this work we impose parametric shapes for the priors for all the parameters of interest
and learn the value of these parameters from a large set of observed coronal loop oscillations.
The ensuing final priors with their shape inferred from the data summarize all the global information
we currently have for the physical properties of coronal loop oscillations.

In the following, we first describe the model used to explain the observations and how
it depends on the physical parameters of interest. Afterwards, we describe the hierarchical
probability model used to explain the complete set of observations and we define the
hierarchical priors used in this work. We also explain how to efficiently sample the
high-dimensional posterior probability distribution function.

\subsection{Coronal loops oscillation generative model}
\label{sec:loop_model}
Observing oscillations in coronal loops with the aim of carrying out magneto-hydrodynamical seismology
is a very difficult task. After an arduous process that requires a detailed analysis of the time evolution of 
images obtained in coronal lines from space missions, the time variation of the 
displacement, $d(t)$, that describes the motion of the coronal loop apex at different time steps,
is obtained \citep[e.g.,][]{aschwanden02}.
In order to extract information from the time evolution of the displacement, this quantity is modeled 
as a combination of a systematic motion of the entire loop and a real oscillatory component. Therefore, the
generative model\footnote{A generative model defines a parametric description of the
signal, taking into account the presence of observational uncertainties and its statistical properties.} for the observations is then:
\begin{equation}
d(t) = d_\mathrm{trend}(t) + d_\mathrm{osc}(t) + \epsilon(t) + b(t),
\label{eq:generative_model}
\end{equation}
where $\epsilon(t)$ represents the uncertainty of the amplitude measurement, while $b(t)$ takes into account
the presence of any remaining uncertainty produced by non-modeled effects like background loops, wrong estimation
of the noise variance, etc. Concerning the standard uncertainty, we assume it has Gaussian statistics, with zero
mean and time-independent variance $\sigma_n^2$. Such a simplification means that measurements at different times are completely
uncorrelated. Additionally, we use an estimation of $\sigma_n$ obtained directly from the observations.
The background component is assumed, for simplicity, to be also Gaussian with
zero mean and time-independent variance $\sigma_b^2$.

The oscillatory component is modeled in terms of a sinusoidal with an exponential
decay as follows:
\begin{equation}
d_\mathrm{osc}(t) = A \sin \left[ \frac{2 \pi}{P} (t-t_0) - \phi_0 \right] \exp \left[- \frac{t-t_0}{\tau_d} \right],
\end{equation}
where $A$ is the amplitude of the oscillatory part, $P$ is its period, $t_0$ is a reference initial time that is
fixed from the observations,
$\phi_0$ is the initial phase of the oscillation and $\tau_d$ is the damping time scale. The detrending
of the oscillatory motion is very difficult to carry out and might depend on an undetermined (potentially large) number
of factors. For this reason, it is customary to use a simple polynomial function, that absorbs all the unknown effects \citep{aschwanden02}:
\begin{equation}
d_\mathrm{trend}(t) = \sum_{i=0}^N a_i (t-t_0)^i,
\end{equation}
where the coefficients $a_i$ are obtained for each coronal loop and the order $N$ is adapted
to the needed complexity. For simplicity, we fix the values of the $a_i$ coefficients to
those obtained by \cite{aschwanden02} because no physical information is extracted from them.

The generative model that we have written does not allow us to extract much physical information. The
period and damping time are purely observational quantities and we need to relate them with
the physical conditions in the loops. To this
end, we propose the resonantly damped MHD kink mode 
interpretation of quickly damped transverse oscillations of coronal
loops \citep{ruderman02,goossens02} to explain the observed period $P$ and damping time $\tau_d$. 
This approximation applies to a straight cylindrically symmetric magnetic
flux tube with a uniform magnetic field pointing along the axis
of the tube. Under the zero plasma-$\beta$ approximation, coronal loops
can be considered to be density enhancements with a constant internal density, $\rho_i$,
a constant external density, $\rho_e < \rho_i$, and a non-uniform transitional layer of thickness 
$l$ that connects both regions. Following \cite{goossens08}, it is possible to give
the following analytical expression for $P$ and $\tau_d$ under the thin tube and thin boundary 
approximations:
\begin{equation}
P = \tau_{A} \sqrt{2} \left( \frac{\xi+1}{\xi} \right)^{1/2} \qquad \mathrm{and} \qquad \frac{\tau_d}{P} = \frac{2}{\pi} \frac{\xi+1}{\xi-1} \frac{1}{l/R}.
\label{eq:period_damping}
\end{equation}
From these considerations, the parameters in which we are interested are the internal Alfv\'en travel 
time, $\tau_{A}$, the density contrast between the tube
and the environment, $\xi=\rho_i/\rho_e$, and the transverse inhomogeneity length scale in units of the radius of the loop, $l/R$.

\subsection{Hierarchy}
\label{sec:hierarchical}
According to the previous model, the oscillatory displacement of the $i$-th coronal loop is determined by the set of
parameters $\thetabold_i=\{\tau_{A},\xi,l/R,A,\phi_0,\sigma_b\}$, where we use the vector $\thetabold_i$
to compact the notation. The Bayesian analysis performed by \cite{arregui_asensio11} demonstrated that
the constraining power of the observations is very limited. Although
$\tau_{A}$ can be successfully estimated from the observations (although with relatively large and
asymmetric error bars), the situation is much worse for the density contrast and the length scale,
with the density contrast being the poorest constrained. \cite{arregui_asensio11} have
shown that the marginal posterior distribution for $\xi$ is very close to the assumed prior distribution,
meaning that there is almost no information in the observations to constrain $\xi$. The reason why, even in
the absence of information for the density contrast, the Alfv\'en travel time can be correctly
recovered has to be found on the specific shapes of the curves in the three-dimensional
space $(\tau_{A},\xi,l/R)$ pertaining to constant values of $P$ and $\tau_d$, as explained
in \cite{arregui07}.

Consider $\Thetabold=\{\thetabold_1,\thetabold_2,\ldots,\thetabold_n\}$ to be a vector of length $6N$ that
contains all the model parameters for all the observed $N$ loops. In a standard Bayesian
approach, the posterior distribution (which encodes the updated information about the
model parameters) is given by:
\begin{equation}
p(\Thetabold| \mathbf{D}) = 
\frac{p(\mathbf{D}|\Thetabold) p(\Thetabold)}{p(\mathbf{D})},
\label{eq:bayes_theorem_simple}
\end{equation}
where $\mathbf{D}=\{D_1,D_2,\ldots,D_n\}$ refers to the observed data, the measured 
time variation of the displacement, $d^\mathrm{obs}(t)$, for all the loops. The function
$p(\mathbf{D}|\Thetabold)$ is the likelihood, that measures the probability of getting a set
of observed displacements for a given combination of the parameters. Viewed as a function
of the parameters $\Thetabold$, the likelihood measures the quality of the parametric model to explain the observations. 
Finally, the function $p(\Thetabold)$ is the prior distribution
that encodes a-priori information about the model parameters, while $p(\mathbf{D})$
is the evidence. Given that $p(\mathbf{D})$ does not depend on the model
parameters, it is just a multiplicative constant and can be dropped from the calculations.

The quantities with physical interest in our problem are $\tau_{A}$, $\xi$ and $l/R$. They are obviously directly unobservable. For
this reason, one cannot use the standard histogram to a set of observed coronal loops with 
the aim of obtaining their general physical properties. It is widely known that Bayesian hierarchical models constitute
a very powerful way to overcome this difficulty (REF). The idea behind hierarchical models is
extremely simple. The priors $p(\Thetabold)$ used in Eq. (\ref{eq:bayes_theorem_simple}) are made
dependent on a set of hyperparameters $\Omegabold$, which are then included in the inference scheme. Formally,
the posterior is given by:
\begin{equation}
p(\Thetabold,\Omegabold| \mathbf{D}) = 
\frac{p(\mathbf{D}|\Thetabold) p(\Thetabold|\Omegabold) p(\Omegabold)}{p(\mathbf{D})},
\label{eq:bayes_theorem_final}
\end{equation}
where we have used the general fact that $p(\Thetabold,\Omegabold)=p(\Thetabold|\Omegabold) p(\Omegabold)$.
Note that we have dropped the dependence of the likelihood on $\Omegabold$, given that $\Omegabold$ are just hyperparameters or,
in other words, parameters of the priors.

If we make the assumption that there is not any correlation between any two
coronal loops from the set of $N$ observations, we can largely simplify Eq. (\ref{eq:bayes_theorem_final}). In such
a case, the likelihood and the priors can be factorized, so that the posterior distribution simplifies to read:
\begin{equation}
p(\Thetabold,\Omegabold| \mathbf{D}) = \frac{1}{p(\mathbf{D})} \prod_{i=1}^N p(D_i|\thetabold_i) p(\thetabold_i|\Omegabold) p(\Omegabold),
\label{eq:final_posterior}
\end{equation}
where we have made use of standard probability calculus. 

Since the global properties of the physical properties are governed by the priors, our aim is to compute
the statistical properties of their parameters, $\Omegabold$. Consequently, and although it might seem
counterintuitive, all the individual physical parameters $\Thetabold$ are nuisance parameters for us
and have to be integrated out from the posterior \citep[e.g.,][]{gregory05}:
\begin{equation}
p(\Omegabold| \mathbf{D}) = \frac{p(\Omegabold)}{p(\mathbf{D})} \prod_{i=1}^N \int d\thetabold_i p(D_i|\thetabold_i) p(\thetabold_i|\Omegabold),
\label{eq:marginal_hyperparameters}
\end{equation}
where we have made used of the fact that the parameters of one loop do not affect those of 
another loop. It is this integration operation the one that propagates information from all individual loops simultaneously
to the hyperparameters.

\begin{figure}
\centering
\includegraphics[width=0.8\columnwidth]{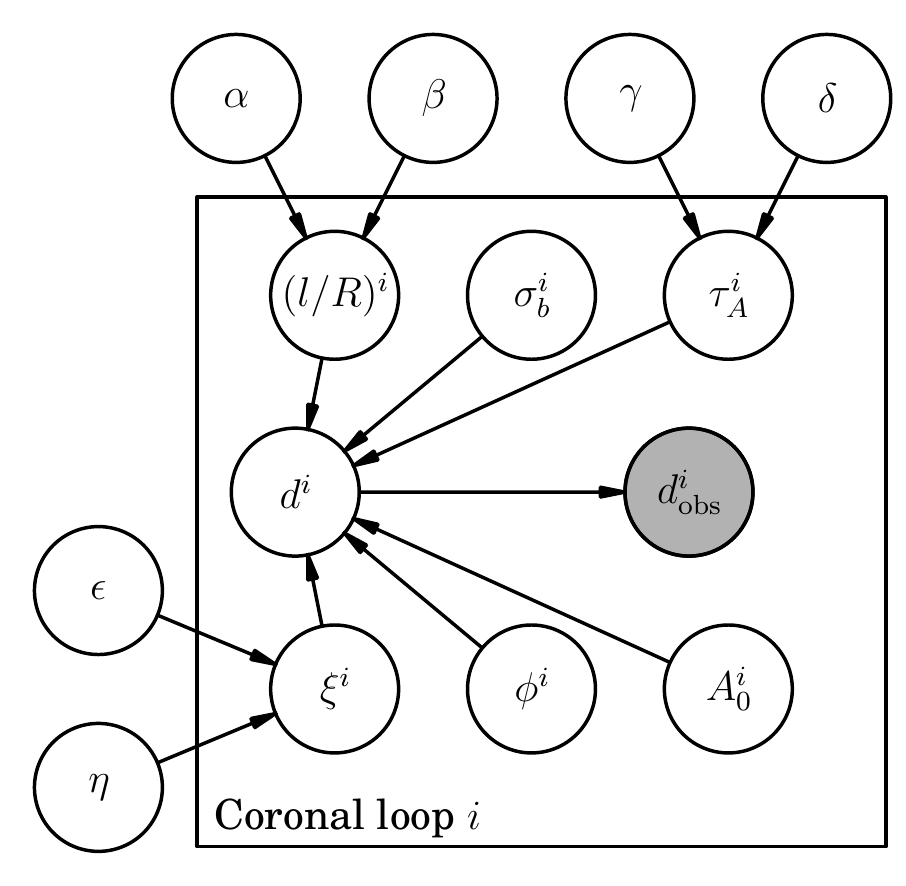}
\caption{Graphical model representing the hierarchical Bayesian scheme that we
used to analyze the set of coronal loop oscillations. Open circles represent random
variables (note that both model parameters and observations are considered as
random variables), while the grey circle represents a measured quantity. The frame
labeled ``Coronal loop $i$'' represents that the model has to be repeated for all
the observations. An arrow between two nodes illustrates dependency. 
The nodes that are outside the frame are the hyperparameters of the
model and are common to all coronal loops.}
\label{fig:graphical_model}
\end{figure}

\subsection{Likelihood}
According to the characteristics of the noise and background components of the generative
model displayed in Eq. (\ref{eq:generative_model}), the likelihood function is a Gaussian. Given that
both $\epsilon(t)$ and $b(t)$ follow the same Gaussian statistics with zero mean although with
different (time-independent) variances, the total likelihood for an individual coronal loop is given by:
\begin{equation}
p(D_i|\thetabold_i) = \mathcal{C} \exp \left[ -\sum_{j=1}^{m_i} \frac{\left(d(t_j)-t_\mathrm{trend}(t_j)-d_\mathrm{osc}(t_j) \right)^2}{2(\sigma_n^2+\sigma_b^2)}\right]
\end{equation}
where $m_i$ is the number of time steps measured for the $i$-th loop and
\begin{equation}
\mathcal{C} = \left(2\pi\right)^{-m_i/2} \left(\sigma_n^2+\sigma_b^2 \right)^{-m_i/2}
\end{equation}

Of importance is to have a good estimation of $\sigma_n$, the variance
of the noise. According to \cite{aschwanden02}, the process of obtaining the
time evolution of the displacement for a given coronal loop is indeed quite complicated. 
For this reason, we take a conservative approach and use $\sigma_n$ equal to 10\% of the maximum absolute displacement 
in each coronal loop. Our results demonstrate that this number is indeed a lower limit to the
actual uncertainty.

\begin{table}
\caption{Priors used in this work}
\label{tab:priors}
\centering
\begin{tabular}{ccc} 
Parameter & Prior & Range\\
\hline
\hline    
$l/R$ & Truncated Gaussian & $[0,2]$ \\
$\xi$ & Shifted inverse Gamma & $[1,\infty)$ \\
$\tau_A$ & Inverse Gamma & $[0,\infty)$ \\
$A$ & Modified Jeffreys' & $[0,\infty)$ \\
$\phi_0$ & Uniform & $[-\pi,\pi]$ \\
$\sigma_b$ & Modified Jeffreys' & $[0,\infty)$ \\
$\alpha$ & Uniform & $[0,\infty)$ \\
$\beta$ & Uniform & $[0,\infty)$ \\
$\gamma$ & Uniform & $[0,\infty)$ \\
$\delta$ & Modified Jeffreys' & $[0,\infty)$ \\
$\epsilon$ & Uniform & $[0,\infty)$ \\
$\eta$ & Modified Jeffreys' & $[0,\infty)$ \\
\hline
\hline\\
\\
\end{tabular}
\end{table}

\subsection{Priors}
\label{sec:priors}
In the hierarchical Bayesian scheme, as important as the definition of the likelihood is the definition of 
suitable priors. As described in the introduction, the idea is that, since the hyperparameters of the priors are 
learnt from all the data \emph{simultaneously}, the resulting prior distributions will be then adapted to the data. As a consequence,
the prior distributions defined hierarchically are generalizations of the
standard calculation of a histogram for quantities that cannot be directly observed, like $l/R$, $\tau_A$ and $\xi$.

To this end, it is favorable to use general probability distributions that naturally fulfill the boundaries for all the parameters. The first
step is to consider the range of variation of the model parameters. After 
\cite{goossens08}, we know that the model parameters have to fulfill
\begin{equation}
l/R \in [0,2], \quad \tau_{A} \geq 0, \quad \xi \gtrsim 1, \quad \phi_0 \in [-\pi,\pi], \quad A \geq 0.
\end{equation}
Additionally, $\tau_{A}$ and $\xi$ have upper boundaries that do not emerge from the theory
but can be estimated based on physical arguments. We use $\tau_{A}^\mathrm{max}=1500$ s and $\xi^\mathrm{max}=100$, although
their precise values are of reduced impact in the final result provided that they are large enough. 

The graphical model describing the hierarchy that we consider in the analysis of
coronal loop oscillations is shown in Fig. \ref{fig:graphical_model}. The selected priors, which
depend on the set of hyperparameters $\Omegabold=\{\alpha,\beta,\gamma,\delta,\epsilon,\eta\}$,
are summarized in Tab. \ref{tab:priors}. We give more details in the following.

\begin{figure}[!t]
\centering
\includegraphics[width=\columnwidth]{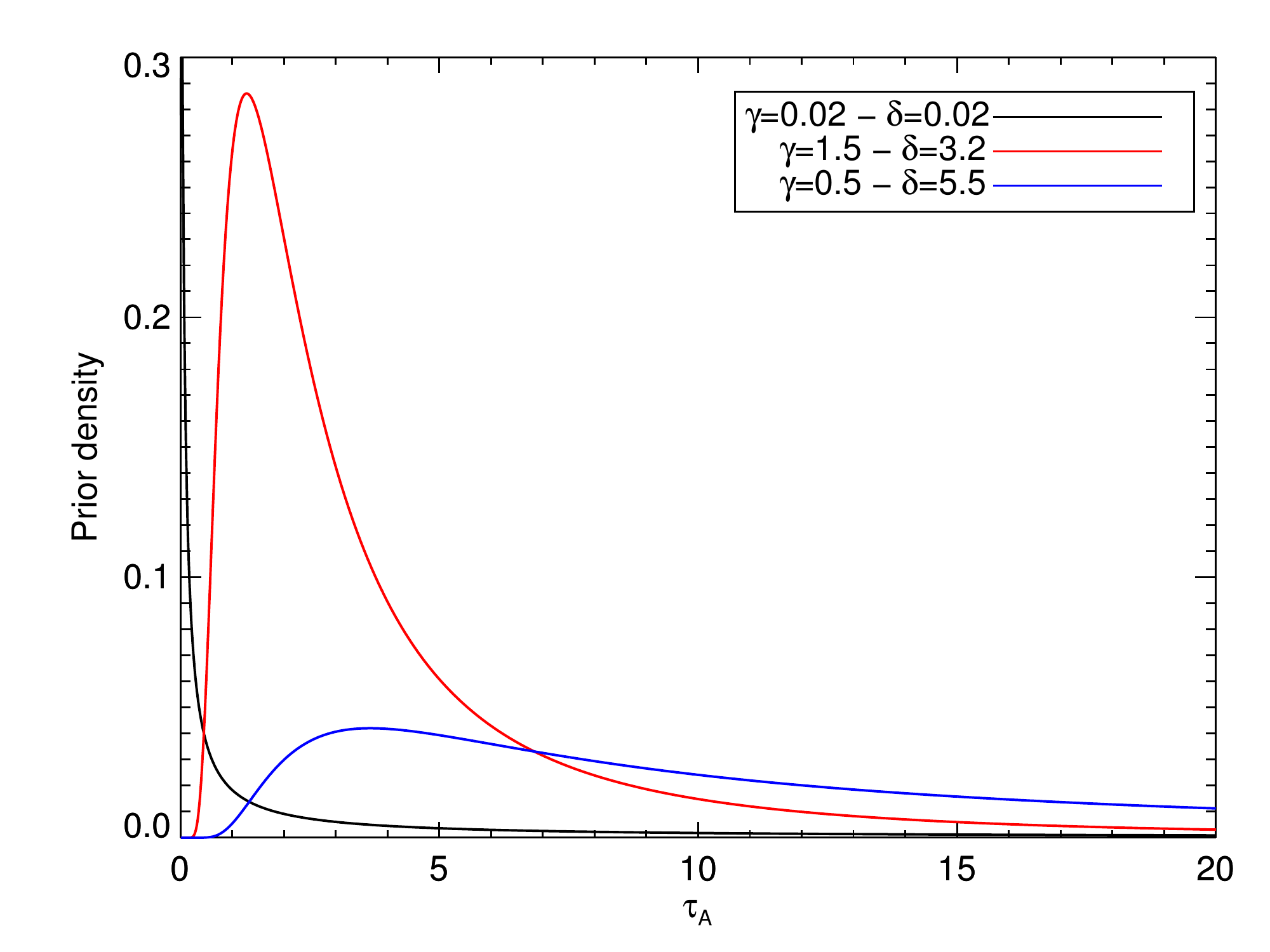}
\caption{Examples of the $\mathrm{IG}(\gamma,\delta)$ distribution, which is a very general distribution
for a positive definite quantity.}
\label{fig:priors}
\end{figure}

\subsubsection{Prior for $l/R$}
The theory says that the transverse inhomogeneity length scale has to lie in the interval $[0,2]$, so it
is advisable to use a prior that automatically fulfills this restriction. We have used a truncated normal
distribution, that depends on two parameters, $\alpha$ and $\beta$, and is given by:
\begin{eqnarray}
\mathrm{TN}(l/R;\alpha,\beta) = 
\left\{ 
\begin{array}{cc}
(\sqrt{2\pi} \beta)^{-1} \exp\left[ -(l/R-\alpha)^2 / (2\beta^2) \right] & 0 \leq l/R \leq 2 \\
0 & \mathrm{otherwise}.
\end{array}
\right.
\end{eqnarray}
Another option that gives very similar results and also depends on two parameters is the scaled Beta prior, defined as
\begin{eqnarray}
\mathrm{Beta}(l/R;\alpha,\beta) = \frac{2^{1-\alpha-\beta}}{B(\alpha,\beta)} (l/R)^{\alpha-1} (2-l/R)^{\beta-1},
\end{eqnarray}
where $B(\alpha,\beta)$ is the beta function \citep[e.g.,][]{abramowitz72}, which 
can be computed in terms of the gamma function as:
\begin{equation}
B(\alpha,\beta) = \frac{\Gamma(\alpha) \Gamma(\beta)}{\Gamma(\alpha+\beta)}.
\end{equation}

\subsubsection{Prior for $\tau_A$}
The Alfv\'en travel time is defined in the interval $[0,\infty)$. A quite general distribution that
is naturally defined in this interval is the inverse gamma distribution, which depends on two
parameters, $\gamma$ and $\delta$:
\begin{eqnarray}
\mathrm{IG}(\tau_A;\gamma,\delta) = \frac{\delta^\gamma}{\Gamma(\gamma)} \tau_A^{-\gamma-1} \exp \left(-\frac{\delta}{\tau_A} \right).
\end{eqnarray}
This distribution has the advantage of describing variables with skewness with only two
parameters. The selection of the inverse gamma distribution is somehow arbitrary and other
distributions like the gamma distribution can be chosen. We have verified with a few of them
that the results are very robust to the precise selection of the functional form, provided
they have sufficient generality. A few examples of the shape of this prior are shown in Fig. \ref{fig:priors}.

\subsubsection{Prior for $\xi$}
The density contrast is a parameter defined in the interval $[1,\infty)$ and scarce information is
available as to what the upper limit can be. For this reason, we choose a shifted inverse gamma
distribution, defined as
\begin{eqnarray}
\mathrm{SIG}(\xi;\epsilon,\eta) = \frac{\eta^\epsilon}{\Gamma(\epsilon)} (\xi-1)^{-\epsilon-1} \exp \left(-\frac{\eta}{\xi-1} \right).
\end{eqnarray}

\subsubsection{Prior for $\sigma_b$}
The standard deviation of the background contribution, $\sigma_b$, is inferred from the data. Given
that it is a scale parameter, it is customary to use a Jeffreys' prior. Given that $\sigma_b$
is defined in the interval $[0,\infty)$ and the Jeffreys' prior is not proper and not well defined
at zero, we propose a modified Jeffreys' prior \citep{gregory05}:
\begin{equation}
MJ(\sigma_b;\sigma_b^0,\sigma_b^\mathrm{max}) = \left[ \left( \sigma_b + \sigma_b^0 \right) \ln \left( \frac{\sigma_b^0+\sigma_b^\mathrm{max}}{\sigma_b^0}\right) \right]^{-1}.
\end{equation}
This prior behaves as a Jeffreys' prior (i.e., as $\sigma_b^{-1}$) for $\sigma_b \gg \sigma_b^0$ and as a 
uniform prior for $\sigma_b \ll \sigma_b^0$. Consequently, the transition parameter $\sigma_b^0$ is a lower boundary
of the Jeffreys' prior. We choose the small value $\sigma_b^0=0.1$. We made sure that this value is
sufficiently small so that the posterior for this parameter peaks at larger values and is therefore
not influenced by its actual value. Concerning $\sigma_b^\mathrm{max}$, it is
made to be very large and its influence on the final results is negligible.

\subsubsection{Prior for $\phi_0$ and $A$}
Without any additional a-priori information, we choose a flat prior for the phase of the oscillation in
the interval $[-\pi,\pi]$. This uniform prior equals $(2\pi)^{-1}$ if $-\pi \leq \phi_0 \leq \pi$ and zero elsewhere.
The amplitude of the oscillation is a scale parameter that is defined in the interval $[0,\infty)$. For this reason, 
we choose a modified Jeffreys' prior:
\begin{equation}
MJ(A;A^0,A^\mathrm{max}) = \left[ \left( A + A^0 \right) \ln \left( \frac{A^0+A^\mathrm{max}}{A^0}\right) \right]^{-1},
\end{equation}
with $A^0=10^{-3}$ (much smaller than the actual amplitude of the oscillation) and a very large $A^\mathrm{max}$.

\subsubsection{Priors for hyperparameters}
The hyperpriors for the hyperparameters $\alpha$, $\beta$, $\gamma$, and $\epsilon$ are all flat in positive real
line. For $\delta$ and $\eta$, given that they can be considered to be scale parameters, we choose modified Jeffreys'
priors with very small transition parameter. However, the final results are very robust and do not depend on the specific hyperpriors.

\subsection{Sampling the posterior}
It is clear that the integrals of Eq. (\ref{eq:marginal_hyperparameters}) cannot be computed analytically.
Therefore, it is necessary to rely on numerical techniques. We carry out the integral using a
technique based on a Markov Chain Monte Carlo \citep[MCMC;][]{metropolis53,neal93}. Instead of the general Metropolis-Hastings
method, we used a Metropolis-within-Gibbs method \citep{tierney94}, which has recently been applied by \cite{sale12}
for mapping the extinction in the Milky Way using a hierarchical Bayesian model\footnote{IDL and Fortran 2003 codes can be downloaded from \texttt{https://github.com/aasensio/mcmc}.}.
The reason for using this scheme is that, in principle, the sampling of the posterior distribution
function for every coronal loop is independent of the others, except for the presence of
the hyperparameters. Therefore, every step of the posterior sampling for each coronal loop can be 
done independently. After one iteration of each chain is carried out, the hyperparameters can be
updated using a standard Metropolis-Hastings rule. This update is then propagated to every
coronal loop. The total length of the converged Markov chains is of the order of a few hundred thousands samples. We verified
that the Markov chains are converged using standard criteria. Finally, the initial 30\% of the chain
is discarded to minimize the sample correlation. As well, we use only one sample every three to further
reduce the correlation.

\begin{figure*}[!t]
\centering
\includegraphics[width=\textwidth]{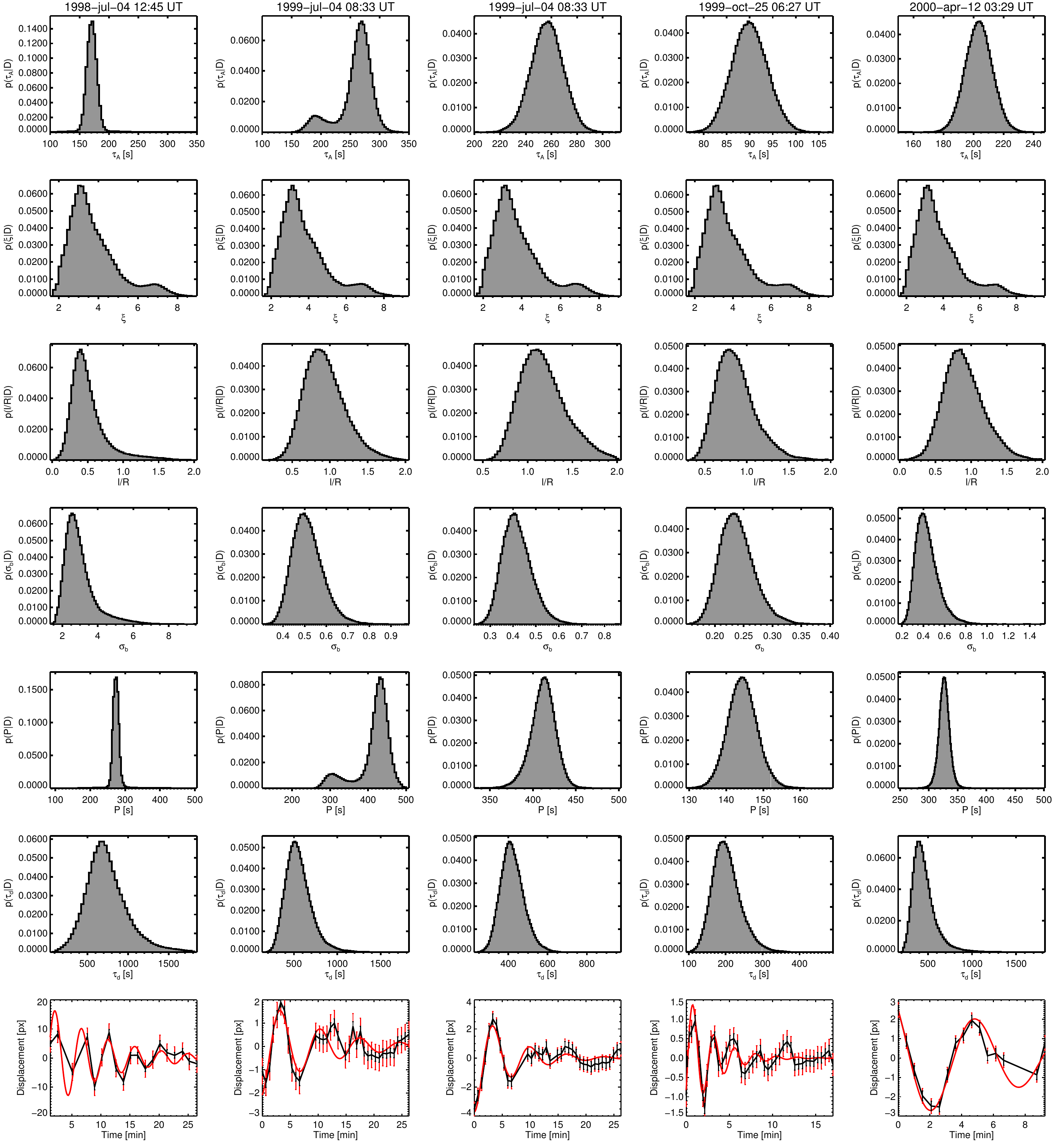}
\caption{Posterior distributions for the model parameters of a sample of five coronal loops. They display the state
of knowledge for all physical parameters of all loops when the observations are taken into account. The inferred
Alfv\'en travel time (first row), density contrast (second row), length scale (third row), the standard deviation
of the background (fourth row), derived oscillation period (fifth row) and damping time (sixth row). The last row shows the original
oscillation corrected for the trend (black curve) and the best fit (red curve). The black error bars are those associated with $\sigma_n$,
while the red error bars are obtained using $\left(\sigma_n^2+\sigma_b^2\right)^{1/2}$.}
\label{fig:posteriors}
\end{figure*}

\begin{figure*}
\centering
\includegraphics[width=\textwidth]{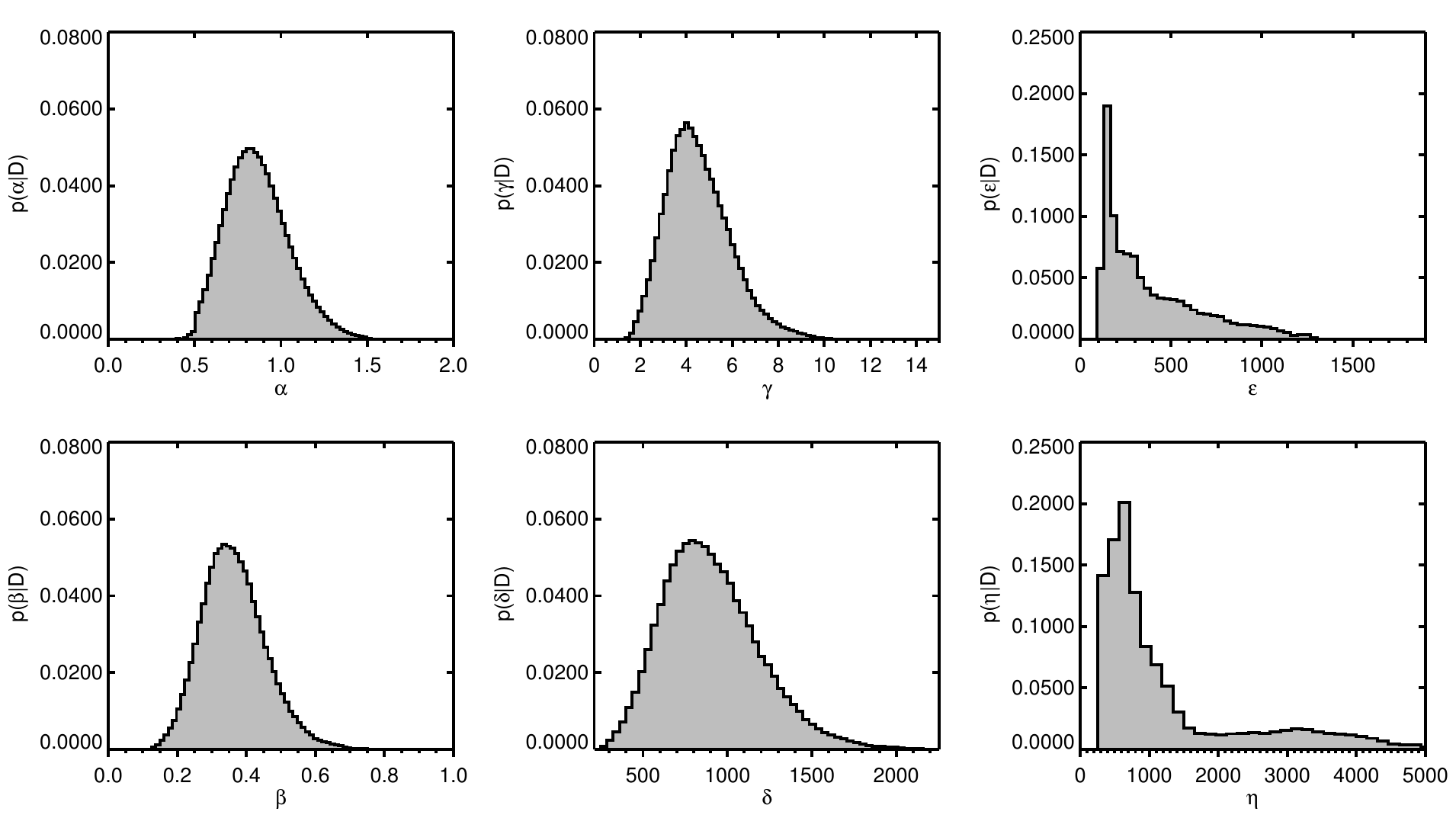}
\caption{Inferred values for the parameters (hyperparameters) that define the assumed probability distribution functions
for $l/R$, $\tau_A$ and $\xi$. The hyperparameters $\alpha$ and $\beta$ define
the prior for $l/R$, $\gamma$ and $\delta$ are used for the prior for $\tau_A$ and $\epsilon$
and $\eta$ define the prior for $\xi$.}
\label{fig:hyperparameters}
\end{figure*}

\subsection{Selection of observations}
Because of the difficulty of observing oscillations in coronal loops, some of 
the curves analyzed by \cite{aschwanden02} do not really display the
behavior that we assume in Sec. \ref{sec:loop_model}. This poses a problem
if the generative model of Eq. (\ref{eq:generative_model}) does not include the term $b(t)$ because
no combination of the model parameters yields a fit to the observations whose
residual is Gaussian with zero mean and variance $\sigma_n^2$. However, the inclusion
of $\sigma_b$ into the inference solves this issue. The observed loops for
which the observation is far from a damped sinusoidal will display a larger $\sigma_b$.

The total number of coronal loops observed by \cite{aschwanden02} is 30. The number of
random variables is then $6N+6=186$, the model parameters for each loop, including the
standard deviation of the background contribution, plus the hyperparameters.

\section{Results}
\label{sec:results}
\subsection{Inference about model parameters}
The output of the MCMC code are samples of the model parameters which are distributed
according to the joint posterior $p(\Thetabold,\Omegabold|\mathbf{D})$. To this, we have
to add the advantage that the Markov chain for a certain parameter is distributed according
to the marginal posterior distribution of this parameter, so the integrals of Eq. (\ref{eq:marginal_hyperparameters})
are automatically obtained. Figure \ref{fig:posteriors} displays the marginal posteriors for a sample of 5
among the 30 coronal loops that we consider in this work. 

The upper row shows the marginal posterior for the Alfv\'en travel time, which are well constrained in all the cases. The marginal
posteriors display a conspicuous peak, although the confidence intervals are clearly asymmetric.
This is similar to the findings of \cite{arregui_asensio11}, although in that paper we did not
fit the whole time evolution. 

The second and third rows show the marginal posteriors for the density contrast and for the length scale. 
It is clear from Eq. (\ref{eq:period_damping}) that the length scale and the density contrast
are intimately related. A fixed value of $\tau_d/P$ can be obtained with an infinite number of combinations
of $\xi$ and $l/R$. Therefore, it is almost impossible to get reliable information for each parameter
separately unless a strong a-priori information is available for any of the two \citep[see][]{arregui07,arregui_asensio11}.
Our results show a very interesting phenomenon that is a direct consequence of the hierarchical scheme. 
The fact that we assume that the priors for $\xi$ and $l/R$ have to be the same for all the observed coronal
loops introduces a large amount of information into the inference. This results into very
well defined posteriors both for the density contrast and the length scale. The strong constraint
imposed by the hierarchical model induces that the density contrast is roughly the same for all 
loops, and the transverse inhomogeneity length scale is the one changing from loop to loop.
We conclude that, under the assumption that the physical properties of all coronal loops are extracted
from common probability distribution functions, the damping time scale is fundamentally determined by the 
transverse inhomogeneity length scale.

The fourth row shows the information inferred for the standard deviation of the background
component. Interestingly, $\sigma_b$ is always non-negligible, meaning that none of the 
observed coronal loops displays a pure damped sinusoidal oscillation. Additionally, the
distribution is very well defined in all cases, so that it is possible to reliably characterize
this background component.

Although $\tau_A$, $\xi$ and $l/R$ are the physical parameters behind the model, it is possible to
compute the marginal posteriors for derived parameters. Using 
Eq. (\ref{eq:period_damping}), we have computed the marginal posteriors for the period and the
damping time, which are shown in the fifth and sixth rows of Fig. \ref{fig:posteriors}. An 
interesting property of these posteriors is that, although some of the model parameters might not 
be strongly constrained, $P$ and $\tau_d$ are very well constrained from the observations.
The marginal posteriors are really close to Gaussian, which reinforces the assumption used in
\cite{arregui_asensio11} of a Gaussian likelihood with diagonal covariance matrix.

Finally, the lowest row of Fig. \ref{fig:posteriors} displays the measured displacement for each loop
and the best fit (roughly equivalent to the least-squares solution, except for the presence of the priors). 
The black error bars are obtained using the estimated value of $\sigma_n$, while the red error
bars are obtained by adding in quadrature $\sigma_n$ and $\sigma_b$.

\begin{figure*}
\centering
\includegraphics[width=\textwidth]{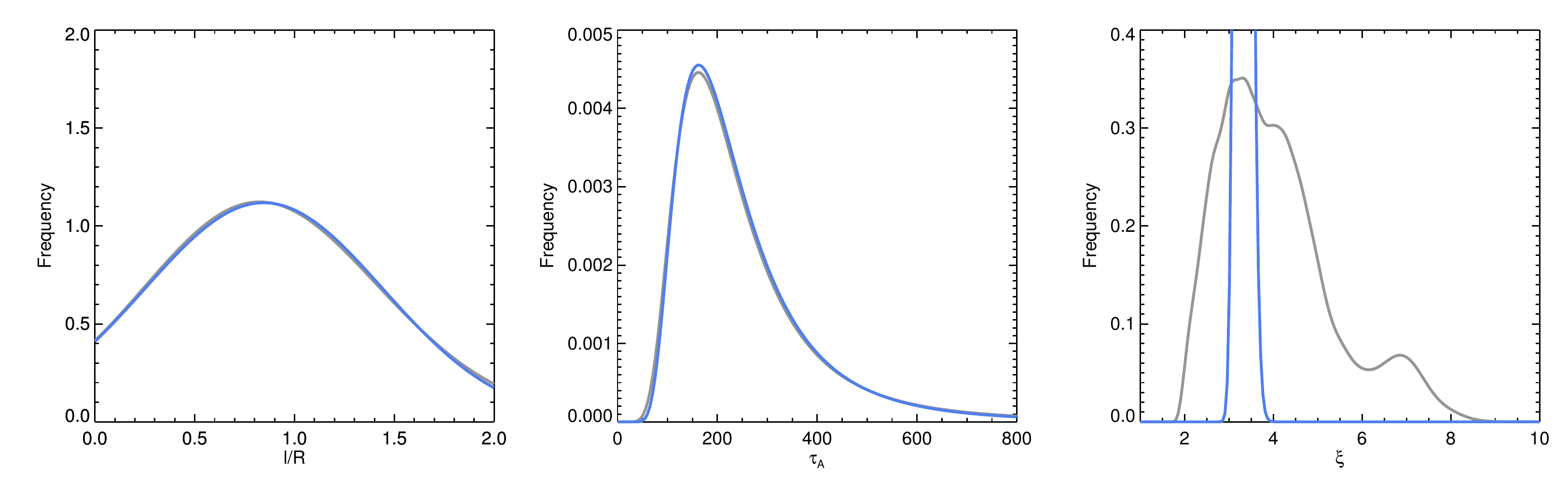}
\caption{Inferred distributions for the transverse inhomogeneity length scale (left panel), the Alfv\'en
travel time (central panel) and the density contrast between the tube and the environment (right panel).
Grey curves represent the marginalized inferred distribution, obtained as the mean of the priors of \S\ref{sec:priors}
with parameters distributed according to Fig. \ref{fig:hyperparameters}. Blue lines are the
distributions of \S\ref{sec:priors} evaluated at the peak of the distributions of Fig. \ref{fig:hyperparameters}.}
\label{fig:hyperpriors}
\end{figure*}

\subsection{Global properties of coronal loops}
The hierarchical structure of our model allows us to obtain the general properties of 
coronal loops. To this end, we show in Fig. \ref{fig:hyperparameters} the inferred distributions
for the hyperparameters that describe the prior distributions described in \S\ref{sec:priors}.
The first column shows the results for $\alpha$ and $\beta$, that are the parameters of the
truncated normal distribution for $l/R$. The results indicate that these hyperparameters have
very well defined values. The median values are $\alpha_\mathrm{med} \approx 0.85$ and
$\beta_\mathrm{med} \approx 0.36$. Likewise, the results for the hyperparameters of the
prior for $\tau_A$ are also well defined, with $\gamma_\mathrm{med} \approx 4.4$ and
$\delta_\mathrm{med} \approx 870$. The situation is less favourable for the hyperparameters
of the prior for $\xi$, probably a consequence of the fact that a single inverse gamma distribution
is not able to capture the complexity of the global properties of $\xi$ over the whole sample
of coronal loops.

Once the hyperparameters are known, it is possible to use this information to get the global properties
of the physical properties of coronal loops. The first approach is to follow what is known as the
type-II maximum likelihood approximation. In this case, we simply evaluate the parametric priors
defined in \S\ref{sec:priors} at the most probable values of their parameters, obtained from the peaks on
Fig. \ref{fig:hyperparameters}. The results are shown as blue lines in Fig. \ref{fig:hyperpriors}.
Another way, that fully takes into account the presence of uncertainties in the hyperparameters, is to use the
$N_s$ Monte Carlo samples of $\alpha$, $\beta$, $\gamma$, $\delta$, $\epsilon$ and $\eta$ from the posterior
to evaluate the following marginalized distributions:
\begin{eqnarray}
\langle p(l/R) \rangle &=& \frac{1}{N_s} \sum_{i=1}^{N_s} \mathrm{TN}(l/R;\alpha_i,\beta_i,0,2)  \nonumber \\
\langle p(\tau_A) \rangle &=& \frac{1}{N_s} \sum_{i=1}^{N_s} \mathrm{IG}(\tau_A;\gamma_i,\delta_i) \nonumber \\
\langle p(\xi) \rangle &=& \frac{1}{N_s} \sum_{i=1}^{N_s} \mathrm{SIG}(\xi;\epsilon_i,\eta_i).
\end{eqnarray}
The previous expressions are the Monte Carlo approximations to the marginalization of the hyperparameters
from the hyperpriors. These distributions are shown as grey lines in Fig. \ref{fig:hyperpriors}.

The distributions shown in Fig. \ref{fig:hyperpriors}, which constitute the main result of this paper, 
represent the underlying distribution from 
which the values of $l/R$, $\tau_A$ and $\xi$ have been sampled, under the assumption that this
global distribution is shared among all the coronal loops. Consequently, they are generalized
histograms of these unobserved quantities, which already take into account any possible degeneracy and 
uncertainty during the inference process. They represent a data-favored updated prior for the parameters 
of the model. These priors can be used in the future when making seismological analysis of coronal loops using the 
resonantly damped magneto-hydrodynamic kink mode interpretation of quickly damped transverse oscillations.

Concerning the transverse inhomogenity length scale, the left panel of 
Fig. \ref{fig:hyperpriors} demonstrates that roughly all allowed values are possible. However, the slight
shift of the distribution shows that there is a small preference for $l/R < 1$. Concerning the Alfv\'en 
travel time, it is clear from the central panel of Fig. \ref{fig:hyperpriors} 
that the most probable value for $\tau_A$ is $\sim 160$ s, with a median value of $\sim 212$ s. The Alfv\'en 
travel time is below $\sim 540$ s are and above $\sim 100$ s with 95\% probability. 
In a surely oversimplified situation in which the typical length $L$ and density $\rho$
of the coronal loop is known with precision, the Alfv\'en travel time limits that
we obtain might be used to put some general constraints on the magnetic field. Given that:
\begin{equation}
\tau_A = \frac{L}{v_A} = \sqrt{\mu_0 \rho} \frac{L}{B},
\end{equation}
where $v_A$ is the Alfv\'en velocity and $B$ is the magnetic field. For instance, if $L \sim 100$ Mm
and $\rho \sim 10^{-14}$ g cm$^{-3}$, we end up with 6 G $\lesssim B \lesssim 35$ G. If the density
is an order of magnitude larger, the magnetic field range increases in a factor $\sqrt{10}$. The most
probable value of the magnetic field, corresponding to the peak of the Alfv\'en travel
time in Fig. \ref{fig:hyperpriors} turns out to be $B \sim 16$ G.
These figures are just plain estimations based on an unrealistic situation in which the properties
of the coronal loop are known.

The information gained for the density contrast is also very interesting. We remind that the
strong constraint for this parameter is a direct consequence of the hierarchical scheme, which
forced the same distribution for all observed coronal loops. According to our results,
the density contrast is above 2.3 and below 6.9 with 95\% probability, with a median value of
3.8.

\begin{figure*}
\centering
\includegraphics[width=0.48\textwidth]{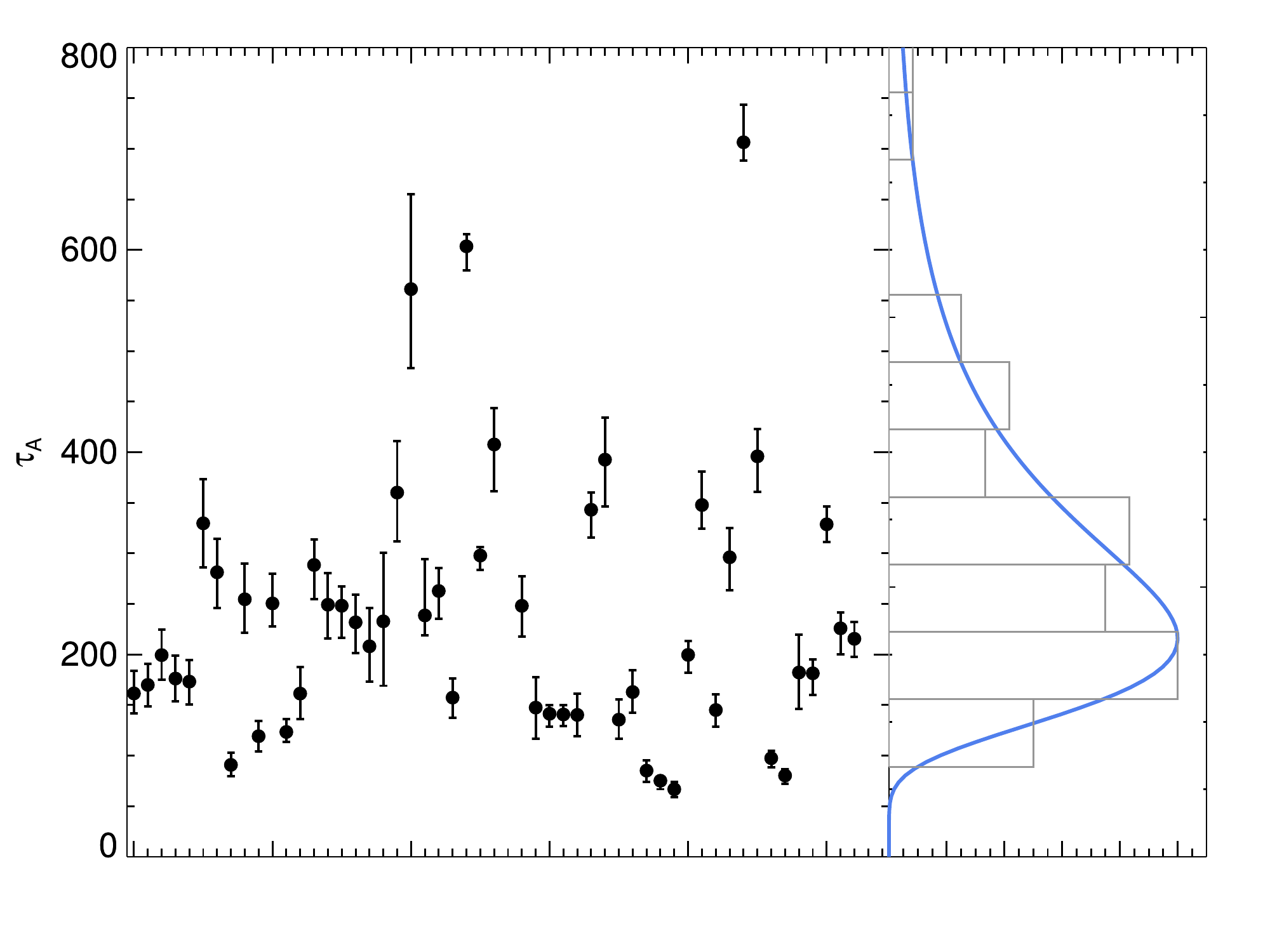}
\includegraphics[width=0.48\textwidth]{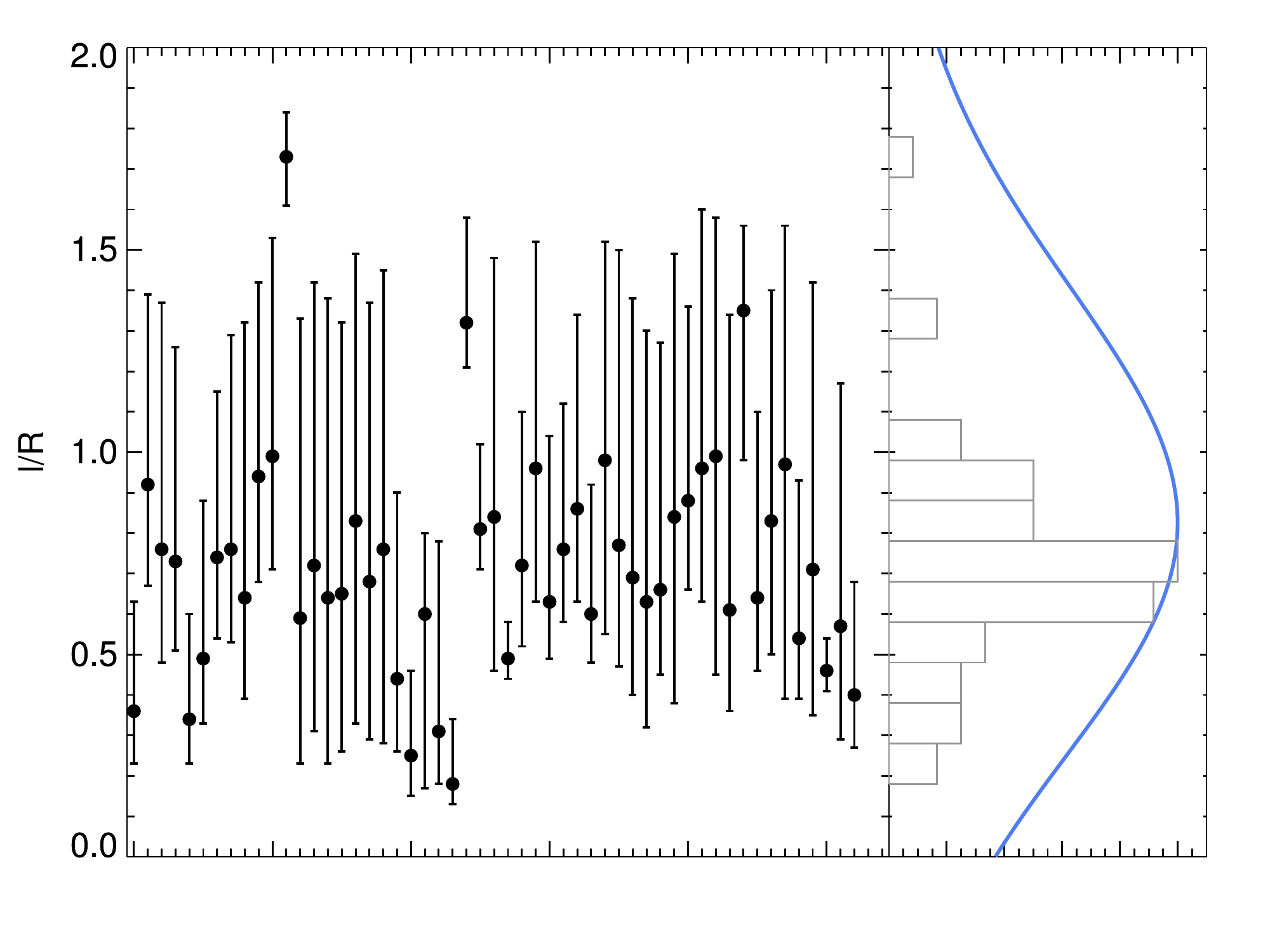}
\caption{Comparison between the inferred distributions shown in Fig. \ref{fig:hyperpriors} and
a simple histogram carried out with the inferred values of $\tau_A$ and $l/R$ using the
formalism of \cite{arregui_asensio11}.}
\label{fig:comparison}
\end{figure*}

Finally, we display in Fig. \ref{fig:comparison} the comparison between our results and what one would
obtain using a simple histogram with the inferred value of the parameters. To this end, we have
used the inferred values of $\tau_A$ and $l/R$ that were obtained by \cite{arregui_asensio11}, 
complemented with the results of applying the Bayesian formalism presented of \cite{arregui_asensio11} 
to the observations collected in Table 1 of \cite{verwichte13}. A Jeffreys' prior in the range
$[1.2,50]$ is used for the density contrast and an uncertainty of 10\% is used for
the period and damping time if no measurement is available.
Although the results are somehow comparable, note
that the error bars are not taken into account in the histogram. This is of special
relevance for $l/R$ and less important for $\tau_A$, where the inferred values are less
uncertain.

\section{Conclusions}
This paper presented the inference of the global physical properties of coronal loops
obtained through MHD seismology. We have obtained the inferred distribution of the Alfv\'en travel
time, size of the transition layer between the surroundings and the coronal loop density enhancement.
These distributions are valid under the assumption that the properties
of all coronal loops are just realizations of some underlying distributions. The results demonstrate 
that sharp transitions between the surrounding and
the internal media are slightly favored. Additionally, we have found that Alfv\'en travel times are in
the interval $[100,540]$ s with 95\% probability. If the length and density of the coronal loop are
known, this poses some constraints on the magnetic field strength in the loop. Likewise, the density
contrast between the loop interior and the surrounding is in the interval $[2.3,6.9]$ with 95\% probability.

Our contribution improves over our previous approach. First, we make the model closer to the
observation, by using a generative model to explain the measured displacements. Second, we
use a method that obtains global information for model parameters that cannot be
directly measured but need to be inferred. Our results allowed us to construct informative priors that
can be used for inversions of individual events. The inference then takes into account
prior beliefs extracted from data.

Apart from the extraordinary difficulty of extracting the oscillations in coronal loops, the potential to massively
apply MHD seismology techniques is now larger than ever thanks to the continuous observations of the Atmospheric Imaging
Assembly \citep[AIA;][]{lemen_aia12} on board the \emph{Solar Dynamics Observatory} (SDO). Its high-temporal
cadence of 12 s and high spatial resolution of $\sim$0.6 arcsec make them the perfect instrument to 
follow these oscillatory events and extract reliable physical information from these coronal events.

\begin{acknowledgements}
We are grateful to Markus J. Aschwanden for kindly providing the measurements of coronal loops oscillations
used in this paper. We thank M. J. Mart\'{\i}nez Gonz\'alez, R. Manso Sainz, M. J. Aschwanden and R. Oliver for
useful suggestions to improve the quality of the manuscript. Financial support by the Spanish Ministry of Economy and Competitiveness 
through projects AYA2010--18029 (Solar Magnetism and Astrophysical Spectropolarimetry) and AYA2011-22846 
is gratefully acknowledged. We also acknowledge financial support through the Ram\'on y
Cajal fellowships and the Consolider-Ingenio 2010 CSD2009-00038 project.
\end{acknowledgements}


\begin{thebibliography}{30}
\expandafter\ifx\csname natexlab\endcsname\relax\def\natexlab#1{#1}\fi

\bibitem[{{Abramowitz} \& {Stegun}(1972)}]{abramowitz72}
{Abramowitz}, M. \& {Stegun}, I.~A. 1972, {Handbook of Mathematical Functions}
  (New York: Dover)

\bibitem[{{Andries} {et~al.}(2005){Andries}, {Arregui}, \& {Goossens}}]{AAG05}
{Andries}, J., {Arregui}, I., \& {Goossens}, M. 2005, \apjl, 624, L57

\bibitem[{{Arregui} {et~al.}(2007){Arregui}, {Andries}, {Van Doorsselaere},
  {Goossens}, \& {Poedts}}]{arregui07}
{Arregui}, I., {Andries}, J., {Van Doorsselaere}, T., {Goossens}, M., \&
  {Poedts}, S. 2007, \aap, 463, 333

\bibitem[{{Arregui} \& {Asensio Ramos}(2011)}]{arregui_asensio11}
{Arregui}, I. \& {Asensio Ramos}, A. 2011, ApJ, 740, 44

\bibitem[{{Arregui} {et~al.}(2013){Arregui}, {Asensio Ramos}, \& {D\'{\i}az,
  A.~J.}}]{Arregui13}
{Arregui}, I., {Asensio Ramos}, A., \& {D\'{\i}az, A.~J.} 2013, ApJ, in press

\bibitem[{{Aschwanden} {et~al.}(2002){Aschwanden}, {de Pontieu}, {Schrijver},
  \& {Title}}]{aschwanden02}
{Aschwanden}, M.~J., {de Pontieu}, B., {Schrijver}, C.~J., \& {Title}, A.~M.
  2002, Sol. Phys., 206, 99

\bibitem[{{Aschwanden} {et~al.}(1999){Aschwanden}, {Fletcher}, {Schrijver}, \&
  {Alexander}}]{aschwanden99}
{Aschwanden}, M.~J., {Fletcher}, L., {Schrijver}, C.~J., \& {Alexander}, D.
  1999, \apj, 520, 880

\bibitem[{{Bovy} {et~al.}(2011){Bovy}, {Hogg}, \&
  {Roweis}}]{bovy_extremedeconv11}
{Bovy}, J., {Hogg}, D.~W., \& {Roweis}, S.~T. 2011, Annals of Applied
  Statistics, 5, 1657

\bibitem[{{Goossens} {et~al.}(2002){Goossens}, {Andries}, \&
  {Aschwanden}}]{goossens02}
{Goossens}, M., {Andries}, J., \& {Aschwanden}, M.~J. 2002, \aap, 394, L39

\bibitem[{{Goossens} {et~al.}(2012{\natexlab{a}}){Goossens}, {Andries},
  {Soler}, {Van Doorsselaere}, {Arregui}, \& {Terradas}}]{goossens12a}
{Goossens}, M., {Andries}, J., {Soler}, R., {et~al.} 2012{\natexlab{a}}, \apj,
  753, 111

\bibitem[{{Goossens} {et~al.}(2008){Goossens}, {Arregui}, {Ballester}, \&
  {Wang}}]{goossens08}
{Goossens}, M., {Arregui}, I., {Ballester}, J.~L., \& {Wang}, T.~J. 2008, A\&A,
  484, 851

\bibitem[{{Goossens} {et~al.}(2011){Goossens}, {Erd{\'e}lyi}, \&
  {Ruderman}}]{goossens11}
{Goossens}, M., {Erd{\'e}lyi}, R., \& {Ruderman}, M.~S. 2011, \ssr, 158, 289

\bibitem[{{Goossens} {et~al.}(2012{\natexlab{b}}){Goossens}, {Soler},
  {Arregui}, \& {Terradas}}]{goossens12b}
{Goossens}, M., {Soler}, R., {Arregui}, I., \& {Terradas}, J.
  2012{\natexlab{b}}, \apj, 760, 98

\bibitem[{{Goossens} {et~al.}(2009){Goossens}, {Terradas}, {Andries},
  {Arregui}, \& {Ballester}}]{goossens09}
{Goossens}, M., {Terradas}, J., {Andries}, J., {Arregui}, I., \& {Ballester},
  J.~L. 2009, \aap, 503, 213

\bibitem[{{Gregory}(2005)}]{gregory05}
{Gregory}, P.~C. 2005, {Bayesian Logical Data Analysis for the Physical
  Sciences} (Cambridge: Cambridge University Press)

\bibitem[{{Lemen} {et~al.}(2012){Lemen}, {Title}, {Akin}, {Boerner}, {Chou},
  {Drake}, {Duncan}, {Edwards}, {Friedlaender}, {Heyman}, {Hurlburt}, {Katz},
  {Kushner}, {Levay}, {Lindgren}, {Mathur}, {McFeaters}, {Mitchell}, {Rehse},
  {Schrijver}, {Springer}, {Stern}, {Tarbell}, {Wuelser}, {Wolfson}, {Yanari},
  {Bookbinder}, {Cheimets}, {Caldwell}, {Deluca}, {Gates}, {Golub}, {Park},
  {Podgorski}, {Bush}, {Scherrer}, {Gummin}, {Smith}, {Auker}, {Jerram},
  {Pool}, {Soufli}, {Windt}, {Beardsley}, {Clapp}, {Lang}, \&
  {Waltham}}]{lemen_aia12}
{Lemen}, J.~R., {Title}, A.~M., {Akin}, D.~J., {et~al.} 2012, Sol. Phys., 275,
  17

\bibitem[{{Metropolis} {et~al.}(1953){Metropolis}, {Rosenbluth}, {Rosenbluth},
  {Teller}, \& {Teller}}]{metropolis53}
{Metropolis}, N., {Rosenbluth}, A.~W., {Rosenbluth}, M.~N., {Teller}, A.~H., \&
  {Teller}, E. 1953, J. Chem. Phys., 21, 1087

\bibitem[{{Nakariakov} \& {Ofman}(2001)}]{Nakariakov01}
{Nakariakov}, V.~M. \& {Ofman}, L. 2001, \aap, 372, L53

\bibitem[{{Nakariakov} {et~al.}(1999){Nakariakov}, {Ofman}, {Deluca},
  {Roberts}, \& {Davila}}]{nakariakov99}
{Nakariakov}, V.~M., {Ofman}, L., {Deluca}, E.~E., {Roberts}, B., \& {Davila},
  J.~M. 1999, Science, 285, 862

\bibitem[{{Neal}(1993)}]{neal93}
{Neal}, R.~M. 1993, {Probabilistic Inference Using Markov Chain Monte Carlo
  Methods} (Dept. of Statistics, University of Toronto: Technical Report No.
  0506)

\bibitem[{{Roberts} {et~al.}(1984){Roberts}, {Edwin}, \& {Benz}}]{roberts84}
{Roberts}, B., {Edwin}, P.~M., \& {Benz}, A.~O. 1984, \apj, 279, 857

\bibitem[{{Ruderman} \& {Erd{\'e}lyi}(2009)}]{ruderman09}
{Ruderman}, M.~S. \& {Erd{\'e}lyi}, R. 2009, \ssr, 149, 199

\bibitem[{{Ruderman} \& {Roberts}(2002)}]{ruderman02}
{Ruderman}, M.~S. \& {Roberts}, B. 2002, ApJ, 577, 475

\bibitem[{{Sale}(2012)}]{sale12}
{Sale}, S.~E. 2012, MNRAS, 427, 2119

\bibitem[{{Schrijver} {et~al.}(2002){Schrijver}, {Aschwanden}, \&
  {Title}}]{schrijver02}
{Schrijver}, C.~J., {Aschwanden}, M.~J., \& {Title}, A.~M. 2002, \solphys, 206,
  69

\bibitem[{{Tierney}(1994)}]{tierney94}
{Tierney}, L. 1994, Annals of Statistics, 22, 2701

\bibitem[{{Uchida}(1970)}]{uchida70}
{Uchida}, Y. 1970, \pasj, 22, 341

\bibitem[{{Verth} {et~al.}(2008){Verth}, {Erd{\'e}lyi}, \& {Jess}}]{Verth08}
{Verth}, G., {Erd{\'e}lyi}, R., \& {Jess}, D.~B. 2008, \apjl, 687, L45

\bibitem[{{Verwichte} {et~al.}(2006){Verwichte}, {Foullon}, \&
  {Nakariakov}}]{verwichte06}
{Verwichte}, E., {Foullon}, C., \& {Nakariakov}, V.~M. 2006, A\&A, 452, 615

\bibitem[{{Verwichte} {et~al.}(2013){Verwichte}, {Van Doorsselaere}, {White},
  \& {Antolin}}]{verwichte13}
{Verwichte}, E., {Van Doorsselaere}, T., {White}, R.~S., \& {Antolin}, P. 2013,
  A\&A, in press

\end{thebibliography}

\end{document}